\documentclass[%
 reprint,
 twocolumn,
 showpacs,
 amsmath,amssymb,
 aps,
 prl,
]{revtex4-1}

\usepackage[usenames,dvipsnames,svgnames,table]{xcolor}
\usepackage{graphicx}
\usepackage[colorlinks=true,
            linkcolor=blue,
            urlcolor=blue,
            citecolor=blue,
            urlbordercolor = white,
            ]{hyperref}
\usepackage[utf8]{inputenc}
\usepackage[english]{babel}

\addto\captionsenglish{}

\def\p{{\partial}}
\def\Re{\mathop{\text{Re}}}
\def\Im{\mathop{\text{Im}}}

\RequirePackage{color}

\begin{document}

\title{Quantized  Laplacian growth, I: Statistical theory of Laplacian growth}
\author{Oleg Alekseev}
\email{teknoanarchy@gmail.com}
%68.05.-n Liquid Liquid., interface
%73.43.-f QH eff
%02.30.Ik Int sys
%47.20.Hw Morph inst
%47.20.Ma, Interfacial. inst
%47.15.km potential flow
%05.	Statistical physics, thermodynamics, and nonlinear dynamical systems (see also 02.50.-r Probability theory, stochastic processes, and statistics)
%05.10.Gg	Stochastic analysis methods (Fokker-Planck, Langevin, etc.)
%67.10.Fj	Quantum statistical theory
%05.40.-a	Fluctuation phenomena, random processes, noise, and Brownian motion (for fluctuations in superconductivity, see 74.40.-n; for statistical theory and fluctuations in nuclear reactions, see 24.60.-k; for fluctuations in plasma, see 52.25.Gj; for nonlinear dynamics and chaos, see 05.45.-a)
\affiliation{%
Chebyshev Laboratory, Department of Mathematics and Mechanics, Saint-Petersburg State University, 14th Line, 29b, 199178, Saint-Petersburg, Russia.
}%
\date{\today}
\begin{abstract}
	We regularize the Laplacian growth problem with zero surface tension by introducing a short-distance cutoff $\hbar$, so that the change of the area of domains is quantized and equals an integer multiple of the area quanta~$\hbar$. The domain can be then considered as an aggregate of tiny particles (area quanta) obeying the Pauli exclusion principle. The statistical theory of Laplacian growth is introduced by using Laughlin's description of the integer quantum Hall effect. The semiclassical evolution of the aggregate is similar to classical deterministic Laplacian growth. However, the quantization procedure generates inevitable fluctuations at the edge of the droplet. The statistical properties of the edge fluctuations are universal and common to that of quantum chaotic systems, which are generally described by Dyson's circular ensembles on symmetric unitary matrices.
\end{abstract}
\maketitle
Laplacian growth is a fundamental problem of pattern formation phenomena in nonequilibrium physics~\cite{Cross93,Gollub99}. It embraces numerous free boundary dynamics including bidirectional solidification, dendritic formation, electrodeposition, dielectric breakdown, bacterial growth, and flows in porous media~\cite{PelceBook}. These processes can be typically represented by diffusion driven motion of an unstable interface between different phases. Despite its seeming simplicity, Laplacian growth raises enormous interest in physics and mathematics, because it produces a variety of universal patterns with remarkable geometrical and dynamical properties~\cite{StanleyBook,BensimonRMP}.

The instability of the interface dynamics is a distinctive feature of Laplacian growth. A relevant example is the viscous fingering in a Hele-Shaw cell, when a less viscous fluid is injected into a more viscous one in a narrow gap between two plates~\cite{BensimonRMP,SaffmanTaylor}. When the effect of surface tension is negligibly small, the interface dynamics becomes highly unstable. This, the so-called \textit{idealized} Laplacian growth problem, possesses a reach integrable structure, unusual for most nonlinear physical phenomena~\cite{Bensimon84,Richardson,WiegmannPRL}.

The idealized problem is ill-defined, because the growing interface typically develops cusplike singularities in a finite time~\cite{Bensimon84}. Thus, a mechanism stabilizing the growth is necessary. For instance, surface tension regularizes the growth by preventing the formation of singularities with an infinite curvature. However, it is a singular perturbation of the system, which ruins a reach mathematical structure of the idealized problem.

In this paper, we regularize Laplacian growth by introducing a short-distance cutoff preventing the cusps production. Namely, we assume that the change of areas of domains is quantized and equals an integer multiple of the area quanta $\hbar$, i.e., the domain is an aggregate of tiny particles (area quanta) obeying the Pauli exclusion principle. The same system also appears in the study of integer quantum Hall effect. Moreover, a semiclassical evolution of the electronic droplet in the quantum Hall regime is known to be similar to Laplacian growth~\cite{WiegmannQH}. Thus, we introduce statistical theory of Laplacian growth by using Laughlin's description of the integer quantum Hall effect.

\textit{Laughlin's theory.} Laughlin's theory correctly approximates ground states of planar droplets of incompressible quantum fluids placed in a (non)uniform background magnetic field~\cite{Laughlin}. Below, we scale the magnetic field $B$ by the uniform field $B_0$, so that  $l_e=\sqrt{2\hbar c/eB_0}$ is a magnetic length, and the magnetic flux is $\Phi=N\Phi_0$, where $\Phi_0=2\pi\hbar c/e$ is a unit of flux quanta. From now on we chose units $e=c=1$, so that $eB_0/c=2$ and $\Phi_0=2\pi\hbar$. Then, Laughlin's wavefunction of $N$ nonintercating electrons in the nonuniform magnetic field reads
\begin{equation}\label{Psi}
	\psi_N(\mathbf{z})=\frac{1}{\sqrt{N! \tau_N}}\prod_{i<j}(z_i-z_j) e^{-\frac{1}{2\hbar}\sum_i\left[z_i\bar z_i-2V(z_i)\right]},
\end{equation}
where $z_j=x_j+ iy_j$ is a position of the $j$th electron, $\mathbf{z}=\{z_1,\dotsc,z_N\}$, the overbar denotes complex conjugation, the normalization factor is
\begin{equation}\label{tau}
	\tau_N=\frac{1}{N!}\int\prod_{i<j}|z_i-z_j|^2\prod_i e^{-\frac{1}{\hbar}\left[z_i\bar z_i-2V(z_i)\right]}d^2z_i,
\end{equation}
and $V(z)$ is the potential of the nonuniform part $\delta B=-\nabla^2 V$ of the total magnetic field $B=B_0+\delta B$. Although  $\delta B=0$ inside the droplet, the nonzero potential $V(z)$ shapes the edge of the droplet---this is a manifestation of the Aharonov-Bohm effect. The potential $V(z)$ is an analytic function inside the droplet, so that,
\begin{equation}\label{V-def}
	V(z)=\Re\sum_{k\geq 1}T_kz^k, \quad T_k=\frac{1}{2\pi k}\int_{\mathbb C} \delta B(z) z^{-k}d^2z,
\end{equation}
where $T_k$ are the multipole moments.

\begin{figure}[t]
\centering
\includegraphics[width=1\columnwidth]{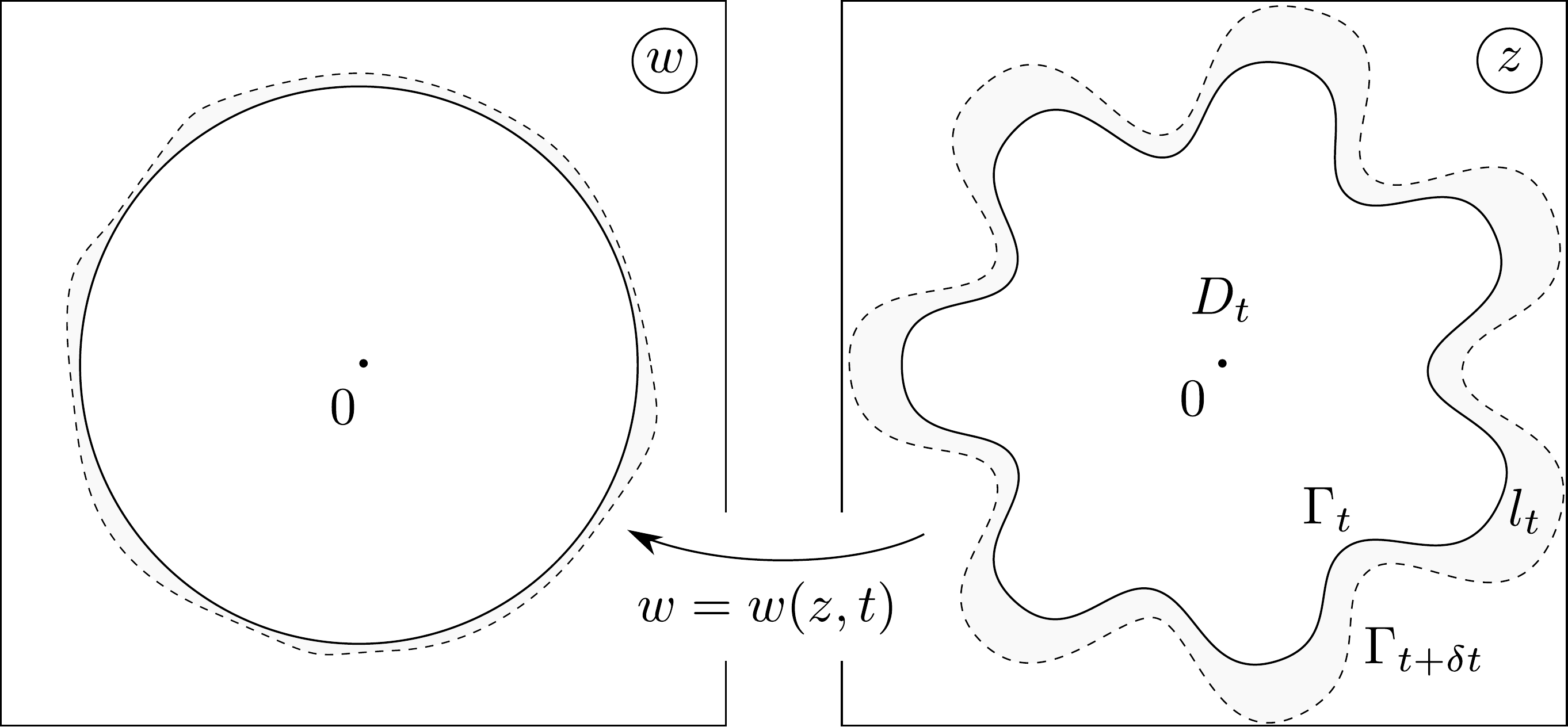}
\caption{\label{map} The time dependent conformal map $w(z,t)$ from $\mathbb C\setminus D_t$ in the $z$ plane to the complement of the unit disk in the $w$ plane. The dashed line in the $z$ plane represents the boundary $\Gamma_{t+\delta t}$ of the advanced domain $D_{t+\delta t}$, which is determined by the density of eigenvalues (the dashed line in the $w$ plane) of the Dyson ensemble. The layer $l_t=D_{t+\delta t}\cap D_t$ is formed by electrons at $\mathbf{z}=\{z_1,\dotsc, z_K\}$, added to the droplet $D_t$ upon increasing the total magnetic flux by $\delta \Phi=2\pi K\hbar$.}
\end{figure}

\textit{Semiclassical limit.} If the number of particles $N$ constituting the droplet is large, whereas their size $\hbar$ is small, and the area of the droplet $\pi t_0=\hbar N$ is kept finite, the aggregate can be described semiclassically. In this limit the density of electrons $\rho_N(z)=\sum_{i=1}^N	\delta^{(2)}(z-z_i)$ is a smooth function of $z$, which steeply drops down at the edge of the droplet~\cite{WZLargeN}. The integral~\eqref{tau} is dominated by the uniform distribution of electrons, which fill the domain $D$ with a constant density. This domain has a sharp boundary (we also assume that the boundary is smooth) and is characterized by the area $\pi t_0=\hbar N$, and the set of external harmonic moments,
\begin{equation}\label{t-def}
	t_k=\frac{1}{\pi k}\int_{\mathbb C\setminus D} z^{-k}d^2z,\quad k=1,2,\dots,
\end{equation}
which equals the multipole moments of the potential~\eqref{V-def}. The semiclassical distribution of electrons also determines the leading contribution to the free energy, $F=\lim_{\hbar\to0}\hbar^2\log \tau_N$~\footnote{By $log$ we denote the logarithm with the base $e$.}, thus establishing a relation with the $\tau$-function of analytic curves~\cite{BOOKtau},
\begin{equation}\label{F-def}
	F(t_0,t_1,\bar t_1,\dotsc)=-\frac{1}{\pi^2}\int_D\int_D\log\left|\frac{1}{z}-\frac{1}{z'}\right|d^2zd^2z'.
\end{equation}
The set of equalities, $t_k=T_k$ ($k=1,2\dotsc$), implies that the system is in mechanical equilibrium, when the electrons are ``frozen'' in equilibrium positions.

\textit{Growth of the droplet.} Below, we assume that system is connected to a large capacitor that maintains a small positive chemical potential slightly above the zero energy of the lowest Landau level. The droplet $D_t$ grows with time $t$ when the total magnetic flux $\Phi$ adiabatically increases~\footnote{We will not indicate the time dependence explicitly if it will not give rise to confusion.}. Adiabatic variations of $V(z)$ deform the shape of the domain (the set of harmonic moments)~\cite{WiegmannQH,AlekseevSM}.

Let us increase the flux by $\delta\Phi=K\Phi_0$, so that the area of the droplet changes by $K\hbar$.  Then, the growth rate is quantized, $Q=Kq$, where $q=\hbar/\delta t$ is the rate quanta.

The semiclassical dynamics of the boundary maximizes the transitional probability between the initial and final states of the quantum system. These states are characterized by Laughlin's wavefunctions $\psi_{N}(\mathbf{z}')$ and $\psi_{N+K}(\mathbf{z}',\mathbf{z})$ respectively. By $\mathbf{z}'=\{z_1',\dotsc,z_N'\}$  we denoted the positions of electrons in the droplet $D_t$, while  $K$ electrons with the coordinates $\mathbf {z}=\{z_1,\dotsc,z_K\}$ form a thin layer, $l_t=D_{t+\delta t}\cap D_t$, with the area $K\hbar$ (see Fig.~\ref{map}). The transitional probability between the droplets, $D_t$ and $D_{t+\delta t}$, is determined by the absolute square of the overlap function~\footnote{The normalization $\int\cdots\int|\psi_{N,K}(\mathbf{z})|^2d^2\mathbf{z}=1$ follows form the properties of the Vandermonde determinant.},
\begin{equation}\label{psi-def}
	\Psi_{N,K}(\mathbf{z})=\int_{\mathbb C}\dotsc\int_{\mathbb C}\overline{\psi_N(\mathbf{z}')}\psi_{N+K}(\mathbf{z}',\mathbf{z})d^2\mathbf{z}'.
\end{equation}
Its absolute square, $|\Psi|^2$, gives a probability of adding $K$ extra particle at the points $\mathbf{z}\in l_t$ of the layer. Using the properties of the Vandermonde determinant, one recasts~\eqref{psi-def} in the form
\begin{multline}\label{psi-tau}
	\Psi_{N,K}(\mathbf{z})=\sqrt{\frac{N!}{(N+K)!}}\frac{e^{-\frac{1}{2\hbar}\sum_i\left[z_i\bar z_i-2V(z_i)\right]}}{\sqrt{\tau_N \tau_{N+K}}}\\
	\prod_{i<j}(z_i-z_j) z_1^{ N}\cdots z_K^{N}e^{-\hbar\sum_{i}D(z_i)}\tau_N,
\end{multline}
and the operator $D(z)=\sum_{k>0}(z^{-k}/k)\p/\p t_k$ generates infinitesimal deformations of the shape of the droplet.

\textit{Semiclassical evolution of the droplet.} Below we only consider the semiclassical limit of the overlap function. In this limit the droplet has a sharp boundary, and the density of electrons is uniform inside the droplet. On expanding~\eqref{psi-tau} in $\hbar$ we recast $|\Psi|^2$ in the form:
\begin{multline}\label{psi2}
	|\Psi_{N,K}(\mathbf{z})|^2\simeq \frac{1}{\sqrt{2\pi^3\hbar}}\prod_{i<j}|z_i-z_j|^{2}\exp\left\{-\frac{2}{\hbar}\sum_i\mathcal A(z_i)\right\}\\
	\exp\left\{-\sum_{i,j}\left[\frac12\p_{t_0}^2-\Re D(z_i)D(z_j)\right]F\right\},
\end{multline}
where $\mathcal A(z)\equiv \mathcal A(z,\bar z)=|z|^2/2-\Re \Omega(z)$ is the modified Schwarz potential written in terms of the potential~\cite{DavisBook},
\begin{equation}\label{Omega}
	\Omega(z)=\sum_{k>0}T_kz^k+t_0\log z-\left[\frac12\p_{t_0}+D(z)\right]F,
\end{equation}
and $F$ is the $\tau$-function of the boundary curve~\eqref{F-def}.

As a function of $z$ the modified Schwarz potential $\mathcal A(z)$ reaches its minimum on the curve $\Gamma=\p D$. The extremum condition, $\bar z=\p_z\Omega(z)$, implies that $\mathcal S(z)=\p_z\Omega(z)$ is the Schwarz function of $\Gamma$. The Schwarz function for a sufficiently smooth curve drawn on the complex plane is an analytic function in a striplike neighborhood the curve, such that $\bar z=\mathcal S(z)$ for $z\in \Gamma$~\cite{DavisBook}.

It is convenient to introduce the (time dependent) conformal map $w(z)$ from the exterior of the droplet $D$ in the physical $z$ plane to the complement of the unit disk is the auxiliary $w$ plane (see Fig.~\ref{map}). The map is normalized in such a manner that $w(\infty)=\infty$ and the conformal radius $r=1/\p_z w(\infty)>0$. A second variation of the free energy~\eqref{F-def} can be then expressed in terms of $w(z)$~\cite{BOOKtau},
\begin{equation}\label{logww}
	\log\frac{w(z_i)-w(z_j)}{z_i-z_j}=\left[-\frac12\p_{t_0}^2+D(z_i)D(z_j)\right]F.
\end{equation}

By using~\eqref{logww} and taking into account that the particles at $\mathbf{z}$ have a continuous distribution (i.e., the density $\rho_K(z)=\hbar\sum_{i=1}^K \delta^{(2)}(z-z_i)$ is a smooth positive function inside the layer~\footnote{In the semiclassical limit $\rho_K(z)$ is a characteristic function of the layer, i.e., $\rho_K(z)=1$ if $z\in l$ and $0$ otherwise.}) we recast~\eqref{psi2} in the form
\begin{multline}\label{psi2-int}
	|\Psi|^2\propto e^{\frac{1}{\hbar^2}\left\{\int_l\int_l\log|w(z)-w(z')|d^2zd^2z'-2\int_l \mathcal A(z)d^2z\right\}}.
\end{multline}

\textit{Stoke's theorem.} Stoke's theorem allows one to reduce the integral over the layer, $l_t$, to the contour integrals along its boundaries, $\Gamma_t$ and $\Gamma_{t+\delta t}$,  which further can be contracted to a single contour integral along $\Gamma_t$. After adding and subtracting the integral $\int_l\int_l\log|z-z'|d^2zd^2z'$ to the exponent of~\eqref{psi2-int}, we obtain
\begin{equation}\label{eq1}
	\begin{aligned}
	&\int_l\int_l\log\left|\frac{w(z)-w(z')}{z-z'}\right|d^2z d^2z'=\\
	&\int_{\Gamma}\int_{\Gamma} \delta h(\zeta)\log\left|\frac{w(\zeta)-w(\zeta')}{\zeta-\zeta'}\right|\delta h(\zeta')|d\zeta||d\zeta'|,
	\end{aligned}
\end{equation}
where $\delta h(\zeta)$ is the width of the layer at $\zeta$, and $\Gamma$ is the inner boundary of the layer $l$, which coincide with the boundary of the domain $D$ (see Fig.~\eqref{map}).

Now, let us apply Stoke's theorem to the integrals,
\begin{equation}\label{G}
	G=\int_l\int_l\log|z-z'|d^2z d^2z'-2\int_l\mathcal A(z)d^2z.
\end{equation}
It is convenient to consider first the following auxiliary integral along the boundary
\begin{equation}\label{I'}
 I(\zeta)=\Re\int_{\Gamma_t}\frac{d\zeta'}{2\pi i}[\mathcal S_{t+\delta t}(\zeta')-\mathcal S_t(\zeta')]\log(\zeta-\zeta'),
\end{equation}
where $\zeta\in l_t$, and $\mathcal S_{t+\delta t}(\zeta)-\mathcal S_t(\zeta)\equiv \delta\mathcal S_t(\zeta)$ is the difference of the Schwarz functions of the curves $\Gamma_{t+\delta t}$ and $\Gamma_t$ correspondingly. The logarithmic branch cut results in the appearance of the term $-\int_{[\zeta,\zeta_0]}\mathcal S_{t+\delta t}(\zeta')d\zeta'$ upon transformation of the contour $\Gamma_t\to \Gamma_{t+\delta t}$ in the first integral in the right hand side of eq.~\eqref{I'}. By $[\zeta,\zeta_0]$ we denoted the path (cut) inside the layer with the endpoints $\zeta\in l_t$ and $\zeta_0\in \Gamma_{t}$. Let us add and subtract the following integral to $I(\zeta)$,
\begin{equation}
	I_\text{cut}(\zeta,\zeta_0)=-\Re\int_{[\zeta,\zeta_0]}\mathcal S_\text{cut}(\zeta')d\zeta',
\end{equation}
where $\mathcal S_\text{cut}(\zeta)$ is the Schwarz function of the cut. Stoke's theorem allows one to recast the sum of the line integrals, $I_{t+\delta t}+I_\text{cut}-I_t$, in the integral over the layer~\footnote{Here, $I_t=(2\pi i)^{-1}\int_{\Gamma_t} d\zeta' \mathcal S_t(\zeta')\log(\zeta-\zeta')$. Note also that $-2\pi i\int_{[\zeta,\zeta]}\mathcal S(\zeta')=(\int_{[\zeta,\zeta_0]^+}-\int_{[\zeta,\zeta_0]^-})\mathcal S(\zeta')\log(\zeta-\zeta')$, where $[\zeta,\zeta_0]^\pm$ are the upper and lower edges of the cut.},
\begin{equation}
	I(\zeta)=\int_{l}\log|\zeta-z'|d^2z'-\Re\int_{[\zeta,\zeta_0]}d\zeta'\left[\mathcal S_t(\zeta')-\mathcal S_\text{cut}(\zeta')\right].
\end{equation}
Since $\int^z \mathcal S(z')dz'=\Omega(z)$ and $\Re \Omega(z)=|z|^2/2$ if $z\in \Gamma$,  the integral $I(\zeta)$ can be rewritten in the form
\begin{equation}
	I(\zeta)=\int_l\log|\zeta-z'|d^2z'-\mathcal A_{t+\delta t}(\zeta),
\end{equation}
where $\mathcal A_t(z)$ is the modified Schwarz potential of $\Gamma_t$. Introducing the integral $I=(2\pi i)^{-1}\int_{\Gamma_t}I(\zeta)\delta\mathcal S_t(\zeta) d\zeta$, and transforming the contour $\Gamma_t\to \Gamma_{t+\delta t}$ one obtains Stoke's theorem for the integrals~\eqref{G},
\begin{equation}\label{G-c}
	G = \int_{\Gamma}\int_{\Gamma} \delta h(\zeta)\log|\zeta-\zeta'|\delta h(\zeta')|d\zeta||d\zeta'|.
\end{equation}
To obtain eq.~\eqref{G-c} we also rewrote the difference of the Schwarz functions in terms of the width of the layer $\delta \mathcal S(\zeta)=2i\sqrt{\mathcal S'(\zeta)}\delta h(\zeta)$ and used $|d\zeta|=\sqrt{\mathcal S'(\zeta)}d\zeta$.

\textit{Statistical theory of Laplacian growth.} The evolution of the droplet $D_t\to D_{t+\delta t}$ is not a deterministic process. The probability of different growth scenarios is given by the absolute square of the overlap function~\eqref{psi2-int}, which is a functional on the grown layer $l_t=D_{t+\delta t}\cap D_t$. Introducing the normal interface velocity $v_n(\zeta, t)$, so that advance of the interface per time $\delta t$ is $\delta h(\zeta)=v_n(\zeta,t)\delta t$, and using eqs.~\eqref{eq1},~\eqref{G-c}, one recasts the transitional probability~\eqref{psi2-int} in the functional on the interface velocity,
\begin{equation}\label{P}
	|\Psi|^2\propto e^{\frac{1}{q^2}\int_{\Gamma}\int_{\Gamma}  v_n(\zeta)\log|w(\zeta)-w(\zeta')|v_n(\zeta')|d\zeta||d\zeta'|},
\end{equation}
where $q=\hbar/\delta t$ is the quanta of the growth rate $Q=Kq$.

The variation of~\eqref{P} shows that $|\Psi|^2$ is maximal when
\begin{equation}\label{v-def}
	v^*_n(\zeta,t)=Q|w'(\zeta,t)|,\quad \zeta\in \Gamma_t.
\end{equation}
This maximum is exponentially sharp when $\hbar\to0$, so all fluctuations around $v_n^*$ are suppressed. Therefore, $v^*_n$ is a classical trajectory for the stochastic growth process. It describes the deterministic Laplacian growth with a sink at infinity. Since $v_n=\Im (\p_t \bar z\p_\phi z)/|\p_\phi z|$, one can rewrite eq.~\eqref{v-def} in the form
\begin{equation}\label{LG}
	\Im\left[\p_t \overline{z(e^{i\phi},t)}\p_\phi z(e^{i\phi},t)\right]=Q,
\end{equation}
which is the classical deterministic Laplacian growth~\cite{BensimonRMP}.

\textit{Stochastic Laplacian growth.} Experimental observations in the Hele-Shaw cell usually show a chaotic dynamics of the interfaces. The mechanism leading to this behavior is missing in the deterministic Laplacian growth problem~\eqref{LG}, where the complex irregular shapes are commonly explained by tiny uncontrollable \textit{initial} deviations of the interface from the perfect shape. Owing to integrability of~\eqref{LG}, a wide class of exact solutions describes in detail a transformation of initially almost perfect interfaces to highly non-trivial irregular shapes~\cite{DMW94}. Although these solutions capture a part of real processes, we do not believe that initial details can encode all complexities of a final shape, as it contradicts to the well-known fact in basic physics that unstable nonlinear systems have, as a rule, a very short memory and quickly ``forget'' initial details.

The nonzero cutoff, $\hbar$, results in the inevitable fluctuations of the interface, which modify the deterministic growth~\eqref{LG}. The analytic treatment of the functional~\eqref{P} can be considerably simplified by mapping it to the auxiliary $w$ plane. Let $\rho(\theta)$ be a macroscopic density function on the unit circle, $|w|=1$, related to the interface velocity in the $z$ plane by the conformal factor,
\begin{equation}\label{v-rho}
	v_n(\zeta,t)=q|w'(\zeta,t)|\rho(\theta), \qquad e^{i \theta}=w(\zeta,t).
\end{equation}
Then, in terms of $ \rho(\theta)$ the functional~\eqref{P} takes the form
\begin{equation}\label{Prho}
	|\Psi|^2\propto \exp\left\{\int_0^{2\pi}\!\!\!\int_0^{2\pi}\rho(\theta)\log|e^{i \theta}-e^{i \theta'}|\rho(\theta')d \theta d \theta'\right\},
\end{equation}
so that the exponent of $|\Psi|^2$ is a negative Coulomb energy of the electric charge distributed on the unit circle with the density $\rho(\theta)$.

The deterministic Laplacian growth~\eqref{LG} is described by the maximum of the functional~\eqref{Prho}, namely, the uniform density $\rho^*(\theta)=K$. The nonuniform density generates fluctuations of the normal interface velocity, so that the growth process is described by the equation
\begin{equation}
	\Im\left[\p_t \overline{z(e^{i\phi},t)}\p_\phi z(e^{i\phi},t)\right]=q \rho(\phi).
\end{equation}

Finally, we note that the distribution function of fluctuations~\eqref{Prho} is similar to Dyson's circular unitary ensemble on symmetric unitary $K\times K$ matrices in the large $K$ limit~\cite{DysonJMP1},
\begin{equation}\label{P-theta}
	|\Psi|^2\propto\prod_{i<j}\left|2\sin\left(\frac{\theta_i-\theta_j}{2}\right)\right|^2,
\end{equation}
so that $\rho(\theta)=\sum_{i=1}^K \delta(\theta-\theta_i)$ is the density of eigenvalues.

\textit{Conclusion.} We regularized Laplacian growth by introducing a short-distance cutoff $\hbar$, which prevents the cusps production. The nonzero cutoff results in the inevitable fluctuations at the interface during the evolution. Because of the instability of Laplacian growth, the system is extremely sensitive to noise: a local destabilization of the interface results in the tip-splitting and formation of fingers of water separated by fjords of oil.

Dyson's ensemble~\eqref{P-theta} (equivalently,~\eqref{P}) describes \textit{static} (time independent) fluctuations of the interface velocity. However, as the fluctuation appears at time $t$, it affects the probabilities of its various values at later time instances. It is expected, that the system will tend to reach the equilibrium state. In future publications we will show that the relaxation to equilibrium in Dyson's ensemble results in the formation of universal patterns with a well developed fjords of oil in the long time asymptotic. Thus, it becomes possible to address \textit{highly nonequlibrium} growth in the framework of \textit{linear} nonequilibrium thermodynamics.

Note that when $K=1$, i.e., only one particle is added to the droplet per time unit, the growth is similar to \textit{diffusion-limited aggregation}~\cite{WittenSanderPRL}. Thus, it is important to obtain Dyson's distribution~\eqref{P-theta} directly from eq.~\eqref{psi2} without the use of the continuum limit. Then, it will become possible to unify both problems, diffusion-limited aggregation and Laplacian growth, within a single framework of (time-dependent) Dyson's circular ensembles.

\bibliography{biblio}{}

%merlin.mbs apsrev4-1.bst 2010-07-25 4.21a (PWD, AO, DPC) hacked
%Control: key (0)
%Control: author (8) initials jnrlst
%Control: editor formatted (1) identically to author
%Control: production of article title (-1) disabled
%Control: page (0) single
%Control: year (1) truncated
%Control: production of eprint (0) enabled
\begin{thebibliography}{23}%
\makeatletter
\providecommand \@ifxundefined [1]{%
 \@ifx{#1\undefined}
}%
\providecommand \@ifnum [1]{%
 \ifnum #1\expandafter \@firstoftwo
 \else \expandafter \@secondoftwo
 \fi
}%
\providecommand \@ifx [1]{%
 \ifx #1\expandafter \@firstoftwo
 \else \expandafter \@secondoftwo
 \fi
}%
\providecommand \natexlab [1]{#1}%
\providecommand \enquote  [1]{``#1''}%
\providecommand \bibnamefont  [1]{#1}%
\providecommand \bibfnamefont [1]{#1}%
\providecommand \citenamefont [1]{#1}%
\providecommand \href@noop [0]{\@secondoftwo}%
\providecommand \href [0]{\begingroup \@sanitize@url \@href}%
\providecommand \@href[1]{\@@startlink{#1}\@@href}%
\providecommand \@@href[1]{\endgroup#1\@@endlink}%
\providecommand \@sanitize@url [0]{\catcode `\\12\catcode `\$12\catcode
  `\&12\catcode `\#12\catcode `\^12\catcode `\_12\catcode `\%12\relax}%
\providecommand \@@startlink[1]{}%
\providecommand \@@endlink[0]{}%
\providecommand \url  [0]{\begingroup\@sanitize@url \@url }%
\providecommand \@url [1]{\endgroup\@href {#1}{\urlprefix }}%
\providecommand \urlprefix  [0]{URL }%
\providecommand \Eprint [0]{\href }%
\providecommand \doibase [0]{http://dx.doi.org/}%
\providecommand \selectlanguage [0]{\@gobble}%
\providecommand \bibinfo  [0]{\@secondoftwo}%
\providecommand \bibfield  [0]{\@secondoftwo}%
\providecommand \translation [1]{[#1]}%
\providecommand \BibitemOpen [0]{}%
\providecommand \bibitemStop [0]{}%
\providecommand \bibitemNoStop [0]{.\EOS\space}%
\providecommand \EOS [0]{\spacefactor3000\relax}%
\providecommand \BibitemShut  [1]{\csname bibitem#1\endcsname}%
\let\auto@bib@innerbib\@empty
%</preamble>
\bibitem [{\citenamefont {Cross}\ and\ \citenamefont
  {Hohenberg}(1993)}]{Cross93}%
  \BibitemOpen
  \bibfield  {author} {\bibinfo {author} {\bibfnamefont {M.~C.}\ \bibnamefont
  {Cross}}\ and\ \bibinfo {author} {\bibfnamefont {P.~C.}\ \bibnamefont
  {Hohenberg}},\ }\href {\doibase 10.1103/RevModPhys.65.851} {\bibfield
  {journal} {\bibinfo  {journal} {Rev. Mod. Phys.}\ }\textbf {\bibinfo {volume}
  {65}},\ \bibinfo {pages} {851} (\bibinfo {year} {1993})}\BibitemShut
  {NoStop}%
\bibitem [{\citenamefont {Gollub}\ and\ \citenamefont
  {Langer}(1999)}]{Gollub99}%
  \BibitemOpen
  \bibfield  {author} {\bibinfo {author} {\bibfnamefont {J.~P.}\ \bibnamefont
  {Gollub}}\ and\ \bibinfo {author} {\bibfnamefont {J.~S.}\ \bibnamefont
  {Langer}},\ }\href {\doibase 10.1103/RevModPhys.71.S396} {\bibfield
  {journal} {\bibinfo  {journal} {Rev. Mod. Phys.}\ }\textbf {\bibinfo {volume}
  {71}},\ \bibinfo {pages} {S396} (\bibinfo {year} {1999})}\BibitemShut
  {NoStop}%
\bibitem [{\citenamefont {Pelce}\ and\ \citenamefont
  {Libchaber}(1988)}]{PelceBook}%
  \BibitemOpen
  \bibfield  {author} {\bibinfo {author} {\bibfnamefont {P.}~\bibnamefont
  {Pelce}}\ and\ \bibinfo {author} {\bibfnamefont {A.}~\bibnamefont
  {Libchaber}},\ }\href@noop {} {\emph {\bibinfo {title} {Dynamics of curved
  fronts}}},\ Perspectives in Physics\ (\bibinfo  {publisher} {Academic
  Press},\ \bibinfo {address} {San Diego},\ \bibinfo {year} {1988})\BibitemShut
  {NoStop}%
\bibitem [{\citenamefont {Stanley}\ and\ \citenamefont
  {Ostrowsky}(1986)}]{StanleyBook}%
  \BibitemOpen
  \bibfield  {author} {\bibinfo {author} {\bibfnamefont {H.~E.}\ \bibnamefont
  {Stanley}}\ and\ \bibinfo {author} {\bibfnamefont {N.}~\bibnamefont
  {Ostrowsky}},\ }\href {\doibase 10.1007/978-94-009-5165-5} {\emph {\bibinfo
  {title} {On Growth and Form}}},\ Nato Science Series E:\ (\bibinfo
  {publisher} {Springer},\ \bibinfo {address} {Netherlands},\ \bibinfo {year}
  {1986})\BibitemShut {NoStop}%
\bibitem [{\citenamefont {Bensimon}\ \emph {et~al.}(1986)\citenamefont
  {Bensimon}, \citenamefont {Kadanoff}, \citenamefont {Liang}, \citenamefont
  {Shraiman},\ and\ \citenamefont {Tang}}]{BensimonRMP}%
  \BibitemOpen
  \bibfield  {author} {\bibinfo {author} {\bibfnamefont {D.}~\bibnamefont
  {Bensimon}}, \bibinfo {author} {\bibfnamefont {L.~P.}\ \bibnamefont
  {Kadanoff}}, \bibinfo {author} {\bibfnamefont {S.}~\bibnamefont {Liang}},
  \bibinfo {author} {\bibfnamefont {B.~I.}\ \bibnamefont {Shraiman}}, \ and\
  \bibinfo {author} {\bibfnamefont {C.}~\bibnamefont {Tang}},\ }\href {\doibase
  10.1103/RevModPhys.58.977} {\bibfield  {journal} {\bibinfo  {journal} {Rev.
  Mod. Phys.}\ }\textbf {\bibinfo {volume} {58}},\ \bibinfo {pages} {977}
  (\bibinfo {year} {1986})}\BibitemShut {NoStop}%
\bibitem [{\citenamefont {Saffman}\ and\ \citenamefont
  {Taylor}(1958)}]{SaffmanTaylor}%
  \BibitemOpen
  \bibfield  {author} {\bibinfo {author} {\bibfnamefont {P.~G.}\ \bibnamefont
  {Saffman}}\ and\ \bibinfo {author} {\bibfnamefont {G.}~\bibnamefont
  {Taylor}},\ }\href {\doibase 10.1098/rspa.1958.0085} {\bibfield  {journal}
  {\bibinfo  {journal} {Proc. R. Soc. London A}\ }\textbf {\bibinfo {volume}
  {245}},\ \bibinfo {pages} {312} (\bibinfo {year} {1958})}\BibitemShut
  {NoStop}%
\bibitem [{\citenamefont {Shraiman}\ and\ \citenamefont
  {Bensimon}(1984)}]{Bensimon84}%
  \BibitemOpen
  \bibfield  {author} {\bibinfo {author} {\bibfnamefont {B.}~\bibnamefont
  {Shraiman}}\ and\ \bibinfo {author} {\bibfnamefont {D.}~\bibnamefont
  {Bensimon}},\ }\href {\doibase 10.1103/PhysRevA.30.2840} {\bibfield
  {journal} {\bibinfo  {journal} {Phys. Rev. A}\ }\textbf {\bibinfo {volume}
  {30}},\ \bibinfo {pages} {2840} (\bibinfo {year} {1984})}\BibitemShut
  {NoStop}%
\bibitem [{\citenamefont {Richardson}(1972)}]{Richardson}%
  \BibitemOpen
  \bibfield  {author} {\bibinfo {author} {\bibfnamefont {S.}~\bibnamefont
  {Richardson}},\ }\href {\doibase 10.1017/S0022112072002551} {\bibfield
  {journal} {\bibinfo  {journal} {J. Fluid Mech.}\ }\textbf {\bibinfo {volume}
  {56}},\ \bibinfo {pages} {609} (\bibinfo {year} {1972})}\BibitemShut
  {NoStop}%
\bibitem [{\citenamefont {Mineev-Weinstein}\ \emph {et~al.}(2000)\citenamefont
  {Mineev-Weinstein}, \citenamefont {Wiegmann},\ and\ \citenamefont
  {Zabrodin}}]{WiegmannPRL}%
  \BibitemOpen
  \bibfield  {author} {\bibinfo {author} {\bibfnamefont {M.}~\bibnamefont
  {Mineev-Weinstein}}, \bibinfo {author} {\bibfnamefont {P.}~\bibnamefont
  {Wiegmann}}, \ and\ \bibinfo {author} {\bibfnamefont {A.}~\bibnamefont
  {Zabrodin}},\ }\href {\doibase 10.1103/PhysRevLett.84.5106} {\bibfield
  {journal} {\bibinfo  {journal} {Phys. Rev. Lett.}\ }\textbf {\bibinfo
  {volume} {84}},\ \bibinfo {pages} {5106} (\bibinfo {year}
  {2000})}\BibitemShut {NoStop}%
\bibitem [{\citenamefont {Agam}\ \emph {et~al.}(2002)\citenamefont {Agam},
  \citenamefont {Bettelheim}, \citenamefont {Wiegmann},\ and\ \citenamefont
  {Zabrodin}}]{WiegmannQH}%
  \BibitemOpen
  \bibfield  {author} {\bibinfo {author} {\bibfnamefont {O.}~\bibnamefont
  {Agam}}, \bibinfo {author} {\bibfnamefont {E.}~\bibnamefont {Bettelheim}},
  \bibinfo {author} {\bibfnamefont {P.}~\bibnamefont {Wiegmann}}, \ and\
  \bibinfo {author} {\bibfnamefont {A.}~\bibnamefont {Zabrodin}},\ }\href
  {\doibase 10.1103/PhysRevLett.88.236801} {\bibfield  {journal} {\bibinfo
  {journal} {Phys. Rev. Lett.}\ }\textbf {\bibinfo {volume} {88}},\ \bibinfo
  {pages} {236801} (\bibinfo {year} {2002})}\BibitemShut {NoStop}%
\bibitem [{\citenamefont {Laughlin}(1990)}]{Laughlin}%
  \BibitemOpen
  \bibfield  {author} {\bibinfo {author} {\bibfnamefont {R.~B.}\ \bibnamefont
  {Laughlin}},\ }in\ \href {\doibase 10.1007/978-1-4612-3350-3} {\emph
  {\bibinfo {booktitle} {The Quantum Hall Effect}}},\ \bibinfo {editor} {edited
  by\ \bibinfo {editor} {\bibfnamefont {R.~E.}\ \bibnamefont {Prange}}\ and\
  \bibinfo {editor} {\bibfnamefont {S.~M.}\ \bibnamefont {Girvin}}}\ (\bibinfo
  {publisher} {Springer},\ \bibinfo {address} {New York},\ \bibinfo {year}
  {1990})\ pp.\ \bibinfo {pages} {233--301}\BibitemShut {NoStop}%
\bibitem [{\citenamefont {Zabrodin}\ and\ \citenamefont
  {Wiegmann}(2006)}]{WZLargeN}%
  \BibitemOpen
  \bibfield  {author} {\bibinfo {author} {\bibfnamefont {A.}~\bibnamefont
  {Zabrodin}}\ and\ \bibinfo {author} {\bibfnamefont {P.}~\bibnamefont
  {Wiegmann}},\ }\href {\doibase 10.1088/0305-4470/39/28/S10} {\bibfield
  {journal} {\bibinfo  {journal} {J. Phys. A}\ }\textbf {\bibinfo {volume}
  {39}},\ \bibinfo {pages} {8933} (\bibinfo {year} {2006})}\BibitemShut
  {NoStop}%
\bibitem [{Note1()}]{Note1}%
  \BibitemOpen
  \bibinfo {note} {By $log$ we denote the logarithm with the base
  $e$.}\BibitemShut {Stop}%
\bibitem [{\citenamefont {Kostov}\ \emph {et~al.}(2001)\citenamefont {Kostov},
  \citenamefont {Krichever}, \citenamefont {Mineev-Weinstein}, \citenamefont
  {Wiegmann},\ and\ \citenamefont {Zabrodin}}]{BOOKtau}%
  \BibitemOpen
  \bibfield  {author} {\bibinfo {author} {\bibfnamefont {I.}~\bibnamefont
  {Kostov}}, \bibinfo {author} {\bibfnamefont {I.}~\bibnamefont {Krichever}},
  \bibinfo {author} {\bibfnamefont {M.}~\bibnamefont {Mineev-Weinstein}},
  \bibinfo {author} {\bibfnamefont {P.}~\bibnamefont {Wiegmann}}, \ and\
  \bibinfo {author} {\bibfnamefont {A.}~\bibnamefont {Zabrodin}},\ }in\
  \href@noop {} {\emph {\bibinfo {booktitle} {Random Matrices and their
  Applications}}},\ Vol.~\bibinfo {volume} {40},\ \bibinfo {editor} {edited by\
  \bibinfo {editor} {\bibfnamefont {P.}~\bibnamefont {Bleher}}\ and\ \bibinfo
  {editor} {\bibfnamefont {A.}~\bibnamefont {Its}}}\ (\bibinfo  {publisher}
  {Cambridge University Press},\ \bibinfo {address} {Cambridge, U.K.},\
  \bibinfo {year} {2001})\ pp.\ \bibinfo {pages} {285--299}\BibitemShut
  {NoStop}%
\bibitem [{Note2()}]{Note2}%
  \BibitemOpen
  \bibinfo {note} {We will not indicate the time dependence explicitly if it
  will not give rise to confusion.}\BibitemShut {Stop}%
\bibitem [{\citenamefont {Alekseev}\ and\ \citenamefont
  {Mineev-Weinstein}(2017)}]{AlekseevSM}%
  \BibitemOpen
  \bibfield  {author} {\bibinfo {author} {\bibfnamefont {O.}~\bibnamefont
  {Alekseev}}\ and\ \bibinfo {author} {\bibfnamefont {M.}~\bibnamefont
  {Mineev-Weinstein}},\ }\href {\doibase 10.1103/PhysRevE.96.010103} {\bibfield
   {journal} {\bibinfo  {journal} {Phys. Rev. E}\ }\textbf {\bibinfo {volume}
  {96}},\ \bibinfo {pages} {010103} (\bibinfo {year} {2017})}\BibitemShut
  {NoStop}%
\bibitem [{Note3()}]{Note3}%
  \BibitemOpen
  \bibinfo {note} {The normalization $\DOTSI \intop \ilimits@ \protect \cdots
  \DOTSI \intop \ilimits@ |\psi _{N,K}(\protect \mathbf {z})|^2d^2\protect
  \mathbf {z}=1$ follows form the properties of the Vandermonde
  determinant.}\BibitemShut {Stop}%
\bibitem [{\citenamefont {Davis}(1974)}]{DavisBook}%
  \BibitemOpen
  \bibfield  {author} {\bibinfo {author} {\bibfnamefont {P.~J.}\ \bibnamefont
  {Davis}},\ }\href@noop {} {\emph {\bibinfo {title} {The Schwarz Function and
  Its Applications}}},\ \bibinfo {series} {The Carus Mathematical Monographs},
  Vol.~\bibinfo {volume} {17}\ (\bibinfo  {publisher} {Mathematical Association
  of America},\ \bibinfo {address} {Buffalo, NY},\ \bibinfo {year}
  {1974})\BibitemShut {NoStop}%
\bibitem [{Note4()}]{Note4}%
  \BibitemOpen
  \bibinfo {note} {In the semiclassical limit $\rho _K(z)$ is a characteristic
  function of the layer, i.e., $\rho _K(z)=1$ if $z\in l$ and $0$
  otherwise.}\BibitemShut {Stop}%
\bibitem [{Note5()}]{Note5}%
  \BibitemOpen
  \bibinfo {note} {Here, $I_t=(2\pi i)^{-1}\DOTSI \intop \ilimits@ _{\Gamma _t}
  d\zeta ' \protect \mathcal S_t(\zeta ')\protect \qopname \relax o{log}(\zeta
  -\zeta ')$. Note also that $-2\pi i\DOTSI \intop \ilimits@ _{[\zeta ,\zeta
  ]}\protect \mathcal S(\zeta ')=(\DOTSI \intop \ilimits@ _{[\zeta ,\zeta
  _0]^+}-\DOTSI \intop \ilimits@ _{[\zeta ,\zeta _0]^-})\protect \mathcal
  S(\zeta ')\protect \qopname \relax o{log}(\zeta -\zeta ')$, where $[\zeta
  ,\zeta _0]^\pm $ are the upper and lower edges of the cut.}\BibitemShut
  {Stop}%
\bibitem [{\citenamefont {Dawson}\ and\ \citenamefont
  {Mineev-Weinstein}(1994)}]{DMW94}%
  \BibitemOpen
  \bibfield  {author} {\bibinfo {author} {\bibfnamefont {S.~P.}\ \bibnamefont
  {Dawson}}\ and\ \bibinfo {author} {\bibfnamefont {M.}~\bibnamefont
  {Mineev-Weinstein}},\ }\href {\doibase
  https://doi.org/10.1016/0167-2789(94)90106-6} {\bibfield  {journal} {\bibinfo
   {journal} {Physica D}\ }\textbf {\bibinfo {volume} {73}},\ \bibinfo {pages}
  {373 } (\bibinfo {year} {1994})}\BibitemShut {NoStop}%
\bibitem [{\citenamefont {Dyson}(1962)}]{DysonJMP1}%
  \BibitemOpen
  \bibfield  {author} {\bibinfo {author} {\bibfnamefont {F.~J.}\ \bibnamefont
  {Dyson}},\ }\href {\doibase 10.1063/1.1703773} {\bibfield  {journal}
  {\bibinfo  {journal} {J. Math. Phys.}\ }\textbf {\bibinfo {volume} {3}},\
  \bibinfo {pages} {140} (\bibinfo {year} {1962})}\BibitemShut {NoStop}%
\bibitem [{\citenamefont {Witten}\ and\ \citenamefont
  {Sander}(1981)}]{WittenSanderPRL}%
  \BibitemOpen
  \bibfield  {author} {\bibinfo {author} {\bibfnamefont {T.~A.}\ \bibnamefont
  {Witten}}\ and\ \bibinfo {author} {\bibfnamefont {L.~M.}\ \bibnamefont
  {Sander}},\ }\href {\doibase 10.1103/PhysRevLett.47.1400} {\bibfield
  {journal} {\bibinfo  {journal} {Phys. Rev. Lett.}\ }\textbf {\bibinfo
  {volume} {47}},\ \bibinfo {pages} {1400} (\bibinfo {year}
  {1981})}\BibitemShut {NoStop}%
\end{thebibliography}%

\end{document}